\documentstyle[12pt]{article}
\begin{document}
 ~
\begin{center}
\begin{Large}
\bf
Exact numerical diagonalization of one-dimensional
interacting electrons
nonadiabatically coupled to phonons
\end{Large}\\
\rm
\bigskip
Gian-Paolo Borghi, Alberto Girlando, and Anna Painelli
\\
\bigskip
Dip. Chimica Generale ed Inorganica, Chimica Analitica e
Chimica Fisica,

Universit\'{a} di Parma, 43100 Parma
(Italy) \\
\vspace{0.5cm}
Johannes Voit \\
\bigskip
Bayreuther Institut f\"{u}r Makromolek\"{u}lforschung
(BIMF) and Theoretische
Physik 1, Universit\"{a}t Bayreuth, D-95440 Bayreuth
(Germany)\footnote{Present Address} \\
and Institut Laue-Langevin, BP 156, F-38042 Grenoble
(France)
\end{center}
\vspace{1cm}

\noindent
We study the role of non-adiabatic
Holstein electron-phonon coupling on the
neutral-ionic phase transition of charge transfer
crystals which can be tuned
 from continuous to discontinuous, using exact numerical 
diagonalization.
The variation of electronic properties through the 
transition is smoothed
by nonadiabaticity. Lattice properties are strongly 
affected, and
we observe both
squeezing and antisqueezing, depending on details of the 
adiabatic
potentials, and identify the quantum uncertainty of the 
phonons as the
most sensitive measure of nonadiabaticity. The adiabatic 
limit is regular
for a continuous transition but
turns out completely
inadequate near a discontinuous transition. The 
relevance of coherent state
approaches is assessed critically.

\vspace{1cm}

\newpage
The adiabatic approximation is one of the most
frequently used approximation
schemes in many-particle physics, especially when 
applied to
electron-phonon coupling. Following Born and Oppenheimer 
(BO) \cite{adia},
one solves the electronic Hamiltonian for fixed nuclear 
coordinates in all
possible configurations $q$. The eigenvalues $E_n(q)$ 
generate the adiabatic
potential which determines the phonon dynamics. The 
associated wave function
is a product of the electronic and phonon functions. The 
anharmonicity of
$E_n(q)$ makes this scheme rather cumbersome, and a more 
practical further
approximation is obtained by expanding to second order 
about the equilibrium
[Herzberg-Teller (HT)
approximation or, in a more solid state language,
mean-field theory augmented by Gaussian
fluctuations].

The usual justification of the BO approximation
is that due to the mass difference of electrons and
ions, the bare
energy scales, Fermi energy ($E_F$) and phonon frequency
($\omega$),
are well separated, with $E_F \gg \omega$. 
However, important electron-electron
interactions
often produce renormalized low-energy excitations
(charge-transfer,
spin-exchange, etc.) whose frequencies may be
comparable
to $\omega$. Also, the polaronic mass
enhancement may give the electrons a mass a sizable
fraction of the ion mass. Moreover, structural phase 
transitions
strongly mix electrons and phonons.
These effects are particularly prominent in one 
dimension (1D)
where the electron-phonon interaction drives the Peierls 
instability
\cite{peierls} and electron-electron interactions are 
strong due to
the narrow bands found in many materials. Is a 
description
based on a product wavefunction, implying decoupling of
the different degrees of freedom, still justified then?

That the adiabatic approximation could fail in 1D was 
suggested by
renormalization group arguments 
\cite{fradkin}-\cite{claude} where the
antiadiabatic ($\omega \rightarrow \infty$) 
limit was identified as the attractive fixed point, and
Quantum Monte Carlo provided some qualitative 
confirmation \cite{fradkin}.
At low $\omega$, however,
the equations became uncontrollable and a reliable
description could only be built on the adiabatic limit
($\omega \rightarrow 0$) \cite{voit,claude}.
Moreover, attention was directed only towards the 
stability of dimerized or
other symmetry-broken ground states.
Another method uses variational
one- and two-phonon coherent states to generate
a lattice displacement and soften the phonon frequency, 
and thereby
increase the quantum lattice fluctuations \cite{zheng}.
However, the product structure of the wave function is 
maintained.
Moreover, the renormalized phonons remain harmonic and, 
in this
sense, one has the antiadiabatic limit of the HT 
approximation.
Despite its current popularity, the
validity of this method has not been assessed 
critically.
Quite detailed
fully nonadiabatic (NA) studies were only possible on 
extremely small
systems.
Several groups could solve two-electron dimers
\cite{dimer},
or a single polaron
on a longer chain \cite{marsig}, or three spinless
electrons
on six sites \cite{carde}. To establish their relevance 
for
a correlated many-particle
system of finite size, a more general framework is 
required.

What is missing to date is an application of the
powerful exact numerical
diagonalization techniques which, despite their
limitation to finite clusters,
have provided so much insight into the physics of
low-dimensional
correlated fermions \cite{dagotto}, to interacting
electrons with \em nonadiabatic \rm
electron-phonon coupling (adiabatic coupling was
frequently considered).
Here we develop such a real-space NA
diagonalization technique, attempting to provide the 
framework called for
above, to identify situations and properties
where the adiabatic approximation breaks
down in 1D, and to critically evaluate the quality of 
the variational
method there \cite{zheng}.

We have chosen as a toy problem a
standard model \cite{nitadia} for the 1D Neutral-Ionic
transition (NIT)
observed in 1D mixed-stack donor--acceptor (DA)
compounds
such as TTF-Chloranil \cite{nitexp}. Our Hamiltonian is
\begin{eqnarray}
\label{hamel}
H_1 & = & -t \sum_{i,s} \left( c^{\dag}_{i,s} c_{i+1,s}
+ {\rm H.c.} \right)
+ U \sum_i n_{i, \uparrow} n_{i, \downarrow} -
\frac{\Delta}{2}
\sum_i (-1)^i n_i \;\;\;, \\
\label{hamphon}
H_2 & = & \frac{1}{2} \sum_i \left( P_i^2 + \omega^2
Q_i^2 \right) \;\;\;
\rightarrow \;\;\; \frac{1}{2} \left( P^2 + \omega^2 Q^2
\right) \;\;\;, \\
\label{hamcoup}
H_{12} & = & g \sqrt{\frac{\omega}{N}} \sum_i Q_i n_i
\;\;\;
\rightarrow \;\;\; g \sqrt{\frac{\omega}{N}} Q \sum_i
(-1)^i
n_i \;\;\;.
\end{eqnarray}
$c_{i,s}^{\dag}$ creates an electron with spin $s$ at
site $i$ which can
hop with a matrix element $t$. $U$ is the on-site
repulsion and $\Delta$
models the energy difference between D and
A sites, but more
complex interactions are possible.
$H_2$ describes local phonon modes with coordinates
$Q_i$, momenta $P_i$, and frequency $\omega$.
There is a local, Holstein-type coupling to the
electrons
with coupling constant $g$. The $N$-site lattice is
half-filled. Neutral D and A sites
are doubly occupied and empty, respectively.
Double ionization of sites is forbidden
in the $\Delta \rightarrow -
\infty$, $U
\rightarrow \infty$ limit; in this limit the only 
relevant
parameter in (1) is $\Delta + U$, the charge transfer 
energy
between adjacent sites \cite{nitadia}.

For small  $(\Delta +U)$ the ground 
state is characterized by small charge transfer from D 
to A,
i.e. by small ionicity $\rho = 1 + N^{-1} \sum_i (-1)^i 
n_i 
\sim 0$.
By increasing $(\Delta +U)$ the system is driven 
to an ionic phase with approximately uniform occupation 
of D 
and A sites ($\rho \sim 1$). No symmetry is broken
at the crossover, and the NIT can be either continuous 
or 
discontinuous. Adiabatically \cite{nitadia}, the NIT is
governed by the
strength of electron-phonon coupling, as measured by the 
small
polaron binding
energy $\varepsilon_{sp} = g^2 / \omega$. 
For large $\varepsilon_{sp}$ the 
NIT is discontinuous; by decreasing $\varepsilon_{sp}$
it crosses a critical point and becomes continuous.
When discontinuous, it is due to an electronic level 
crossing
(two-well transition);
when continuous, the level crossing is avoided, and 
there is a simple
crossover (single-well transition).

We limit ourselves to the $q=2k_F = \pi$-mode (in the
extended zone representation) \cite{nitadia}, and the
corresponding forms of $H_2$ and $H_{12}$ are indicated
in Eqs.~(\ref{hamphon}) and (\ref{hamcoup}).
Other modes are not expected to contribute
in any significant manner because (i) the Peierls
divergence in this commensurate system singles
out the $2k_F = \pi$-mode; (ii) it is further enhanced
in half-filled bands
due to Umklapp scattering; (iii) the external potential
$\Delta$ generated
by the donor-acceptor alternation with the same
wavevector,
quenches the phonons into this preselected mode.
In other words, the NIT involves relaxation of 
just this mode -adiabatically the corresponding
coordinate is proportional to the order parameter
$\rho$- and a comparison of exact and adiabatic results 
obtained in this approximation gives important 
information
on NA effects near a structural phase transition.
In general, however, this wavevector is selected 
dynamically;
this may  renormalize
various properties and has to be critically examined.

The Hilbert space of a phonon is of infinite dimension. 
We
truncate the basis by adopting a convergence
criterion on the energies of the
low-lying states. The number of phonons ($n_{\rm max}$)
required for convergence
strongly depends on the phonon representation used. An 
initial attempt to use
a Bargmann representation for the reference
phonons was limited by the large $n_{\rm max}$
needed to generate the adiabatic lattice
displacement in the neutral phase (reference phonons are 
defined
with respect to the $\rho = 1$ state) \cite{tueb}.
In order to separate the
adiabatic contribution we proposed a two step procedure 
\cite{seoul}.
In the first step, we generate a
set of adiabatic HT basis functions. We determine from 
an adiabatic
diagonalization the equilibrium position $Q_0$, and 
softened
frequency $\Omega = \omega \sqrt{1 - \chi_{\rm el}(\pi) 
\varepsilon_{sp}}$
of the HT harmonic oscillator ($\chi_{\rm el}$ is the 
electronic
susceptibility at the mean-field $Q_0$). In a second 
step, the
full Hamiltonian is represented in HT basis, forming
a band diagonal symmetric matrix, and diagonalized.
The choice of the optimal harmonic adiabatic basis 
reduces
the number of phonon states required for convergence
by a factor $\sim 6$ far from the NIT and $\sim 3$ close 
to the
NIT where nonadiabaticity is important.
This is an important achievement because the
dimension of the NA basis increases as $n_{\rm max}^m$ 
when
$m$ phonon modes are included. The choice of the HT 
basis is critical in
allowing calculations for larger systems and/or more 
than one phonon mode.

Calculations have been performed for rings of $N = 4
\ldots {12}$ sites on a 32 Mbyte DEC 3000 AXP/400 
workstation.
We display  results only for 10-site rings.  
In order to keep the curves in the same parameter
range the results are given against
 $\gamma = (\Delta + U - \varepsilon_{sp})/2$; 
energies are measured in $\sqrt{2}t$ units. A systematic
finite size analysis to the thermodynamic limit will be
reported in a future publication.
The HT results depend on the phonons only through
$\varepsilon_{sp}$. We present results for two different
values of $\varepsilon_{sp}$. For weak coupling,
$\varepsilon_{sp} = 1.28$, the NIT is continuous in the 
adiabatic limit,
whereas for strong coupling ($\varepsilon_{sp}= 2.56$) 
it is
discontinuous. This is gauged through the 
$\rho(\gamma)$-curves which in
the first case exhibit a continuous increase from $\rho 
\ll 1$ at $\gamma
\leq 0$ to $\rho \sim 1$ at $\gamma \geq 0.5$ while 
there is a finite
jump at $\gamma_c$ in the second case ($\gamma_c \sim 
0.2$ has a weak
residual dependence on $\varepsilon_{sp}$) 
\cite{nitadia}.
In both cases, due to phonon
quantum fluctuations, the NA $\rho(\gamma)$
curves are smoother than the HT ones, in agreement with 
expectation.
In particular, within the resolution of our data and for 
the parameter
values used in the figures below, the adiabatically
discontinuous
transitions seem to become continuous.

A particularly pronounced influence of nonadiabaticity
is found in the properties of
the phonon subsystem. Figure 1 displays the phonon
occupation number
in the ground state. The HT ground state is
the phonon vacuum. A finite occupation is generated in
the vicinity of
the NIT by mixing in higher states from the
HT solution. This effect is an
order of magnitude stronger for the two-well
transition. However, in this case the deviations
increase with decreasing phonon frequency which is 
counterintuitive.
On the other hand, the adiabatic potentials at the BO
level are clearly asymmetric.
It is then necessary to
discriminate between truly nonadiabatic and (adiabatic) 
anharmonic effects
(the anharmonic oscillator has a $n > 0$ in the ground 
state).
To this end, we have performed  adiabatic BO 
calculations.
For the single-well transition, the deviations from HT 
are very small:
not larger than 0.02, and decreasing with $\omega$.
For the two-well transition at large frequency the BO 
and NA
curves are similar, but the BO deviations from HT 
decrease with decreasing
$\omega$. Therefore, HT represents the low-frequency 
limit of adiabatic
(BO) models for both single- and two-well  transitions 
(i.e.
adiabatic anharmonic effects vanish in the $\omega 
\rightarrow 0$ limit),
but the adiabatic limit itself is well-defined only for 
single-well potential.
For two-well transitions  the $\omega \rightarrow 0$ 
limit
of the true NA system does differ from the adiabatic 
limit.
The point is that
the BO approximation is an expansion about a 
well-defined
equilibrium position \cite{adia} and cannot be applied 
consistently
to an adiabatic potential with several degenerate 
minima. Of course, a
Hamiltonian composed of such a potential and a kinetic 
term can be solved,
but it is not guaranteed that it represents an adiabatic 
approximation to
the original electron-phonon problem. In this case, the 
fully NA calculation
is clearly preferrable.

In Fig. 2 we show the fluctuations of the phonon 
coordinate
($\delta Q^2$). Since the HT phonon is
softened with respect to the bare phonon, the
fluctuations of the HT coordinate are larger than those
of the reference phonon coordinate. This 
softening-induced
``squeezing'' of the phonon states is rather obvious and 
is included through
two-phonon coherent states in Zheng's approach 
\cite{zheng}.
From the curves shown in Fig. 2 it turns out however 
that
NA fluctuations can either squeeze or antisqueeze phonon
states with respect to HT phonons. Which effect occurs 
depends
on details of the adiabatic potentials and cannot be 
decided a priori.
For the single-well transition, 
the excited state potential is narrower than
the ground state one, and the NA mixing therefore 
reduces
$\delta Q^2$.
For the two-well transition, the adiabatic
potential has a more
extreme minimum in the excited than in the ground state,
allowing additional
excursions of the phonon coordinate and thus increasing
$\delta Q^2$ with respect to HT. We stress however that 
the
squeezing or antisqueezing shown in Fig.~2 cannot be
related to a softening or hardening of the
phonons and therefore cannot be transformed away by 
canonical transformations.
In fact $\delta Q^2$ is related to a frequency only for 
the
harmonic oscillator, where the coordinate  fluctuations
are in inverse relation with the momentum fluctuations.

In Fig.~3 we show the behavior of
$\sqrt{\delta P^2 \delta Q^2}$. The ground state of a
harmonic oscillator or any coherent phonon state
has minimal uncertainty: $\sqrt{\delta P^2 \delta Q^2} = 
1/2$.
Of course, $\delta P^2$ peaks in the sense opposite to 
$\delta Q^2$ but
importantly, the fully NA $\sqrt{\delta P^2 \delta Q^2}$ 
is increased
from the coherent state picture by about
10\% near the single-well and about 90\% near the
two-well transition.
Even more interesting is the comparison with
$\sqrt{\delta P^2 \delta Q^2}$ obtained in BO:
its deviations from the minimum (1/2) value are 
completely negligible
for the single-well and (with all the \em caveats \rm 
above)
not larger than 10\% for the two-well
transition. The important implication is that (i) the 
anharmonicity of the
BO potential cannot effectively account for the large 
incoherence
generated by NA fluctuations, and (ii) $\sqrt{\delta P^2
\delta Q^2}$ is a sensitive probe of precisely the NA
effects. Therefore, at least for the phonon subsystem, 
any
picture based on coherent phonon states (like the
two-phonon squeezed states) is inadequate near a 
(especially
multi-well) phase transition.

Similar results are found when intersite
electron-electron interactions are included.
 Unlike above, data not shown here seem
to indicate that for large enough interactions 
the NA transition remains discontinuous up to some 
finite
critical phonon frequency before turning continuous.

In summary, we have exactly solved a NA
electron-phonon problem with real space numerical
diagonalization. Important progress was made by using a 
renormalized
adiabatic Herzberg-Teller basis which strongly reduces 
the number of
phonon states required for convergence. In the future, 
this will
allow the treatment of larger systems and/or inclusion 
of more phonon modes.
(i) In agreement with expectation, we find that 
nonadiabatic electron-phonon
coupling smoothes the variation of electronic properties 
through the
Neutral-Ionic transition, and that sizable corrections 
to the lattice
properties are generated there. (ii) We find that 
nonadiabaticity can either
squeeze or antisqueeze phonon states, depending on 
details of the adiabatic
potentials while current applications of coherent states 
to similar problems
only allow for squeezing. (iii) By
comparing with Born-Oppenheimer and coherent-state 
calculations, we
identify $\sqrt{\delta P^2 \delta Q^2} - 1/2$ as a 
quantity
particularly sensitive to nonadiabaticity and the 
concomitant loss of
coherence of the phonons while it is insensitive to the 
anharmonic nature of
the adiabatic potential. This loss of coherence is a 
universal feature
of nonadiabaticity and must therefore be important near 
any structural
phase transition, irrespective of its details.
The coherent state picture is bound
to miss the incoherence generated by the nonadiabatic 
mixing of electronic
and phonon degrees of freedom. It would be interesting 
to see if a
variational description of such a situation could
be achieved by introducing another parameter varying 
$\delta P^2$ and
$\delta Q^2$ independently. (iv) Moreover, fully NA 
solutions
are generally preferrable to an anharmonic 
Born-Oppenheimer calculation despite
the bigger matrices involved. While for single-minimum 
potentials, they converge
against BO as $\omega \rightarrow 0$, the NA calculation 
avoids the consistency
problems of BO when several (nearly) degenerate minima 
emerge in the adiabatic
potential.

Some (speculative) predictions can be made about the 
behaviour of systems which
undergo symmetry-breaking structural transitions 
\cite{peierls}. Here the
lower adiabatic potential will change from 
single-minimum to a
degenerate double minimum
at the transition, while the first excited one in 
general has a narrower
single-minimum shape. One will run into the same 
consistency problems
of the BO approximation below that transition as in our 
two-well case.
On the other hand, we expect
that the NA coupling to the first electronic excited 
state will reduce the
phonon coordinate fluctuations and therefore produce 
antisqueezing.

We acknowledge useful discussions with Zoltan Soos and
Jean Bellissard,
the hospitality of the Institute for Scientific
Interchange (Torino) and support from DFG via SFB 
279/B4.

\newpage
\noindent
{\bf Figure Captions} \\
{\bf Figure 1:}
Phonon occupation number $n$
of the ground state as a function of 
$\gamma = (\Delta + U - \varepsilon_{sp})/2$. In the
adiabatic Herzberg-Teller approximation $n \equiv 0$.
Left panel: dash-dotted line: $\omega = 0.125$
dashed line: $\omega = 0.5$,
full line: $\omega = 2.0$. Right panel: dashed line: 
$\omega = 0.5$,
full line: $\omega = 1.0$.

\noindent
{\bf Figure 2:}
Fluctuations $\delta Q^2$ of the phonon coordinate.
Adiabatically $\delta Q^2 = 1$ in units of the 
renormalized
frequency $\Omega$. Legend as in Fig. 1.

\noindent
{\bf Figure 3:}
Uncertainty of phonon states. Adiabatically
$\sqrt{\delta P^2 \delta Q^2} = 1/2$. Legend as in Fig. 
1.

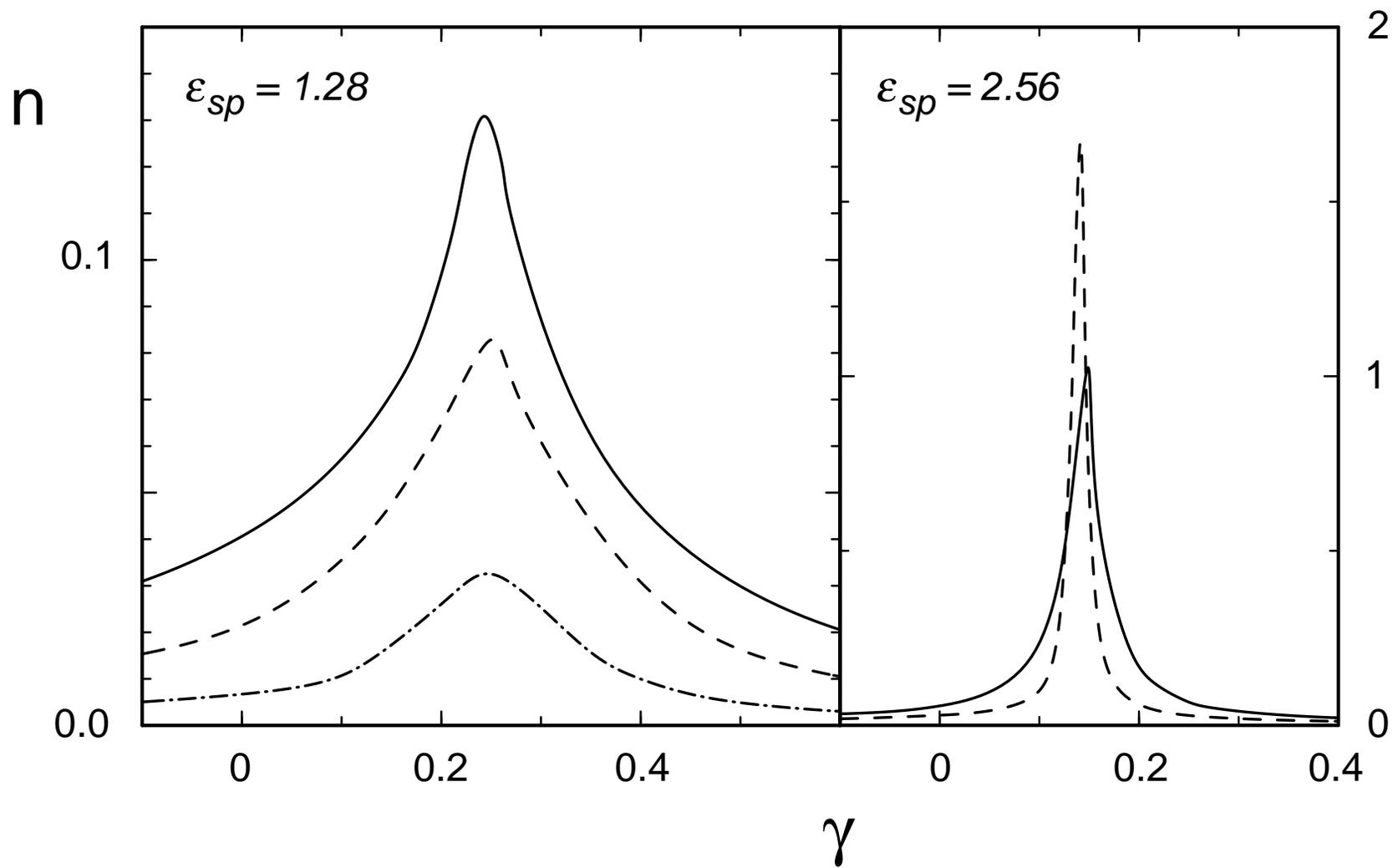
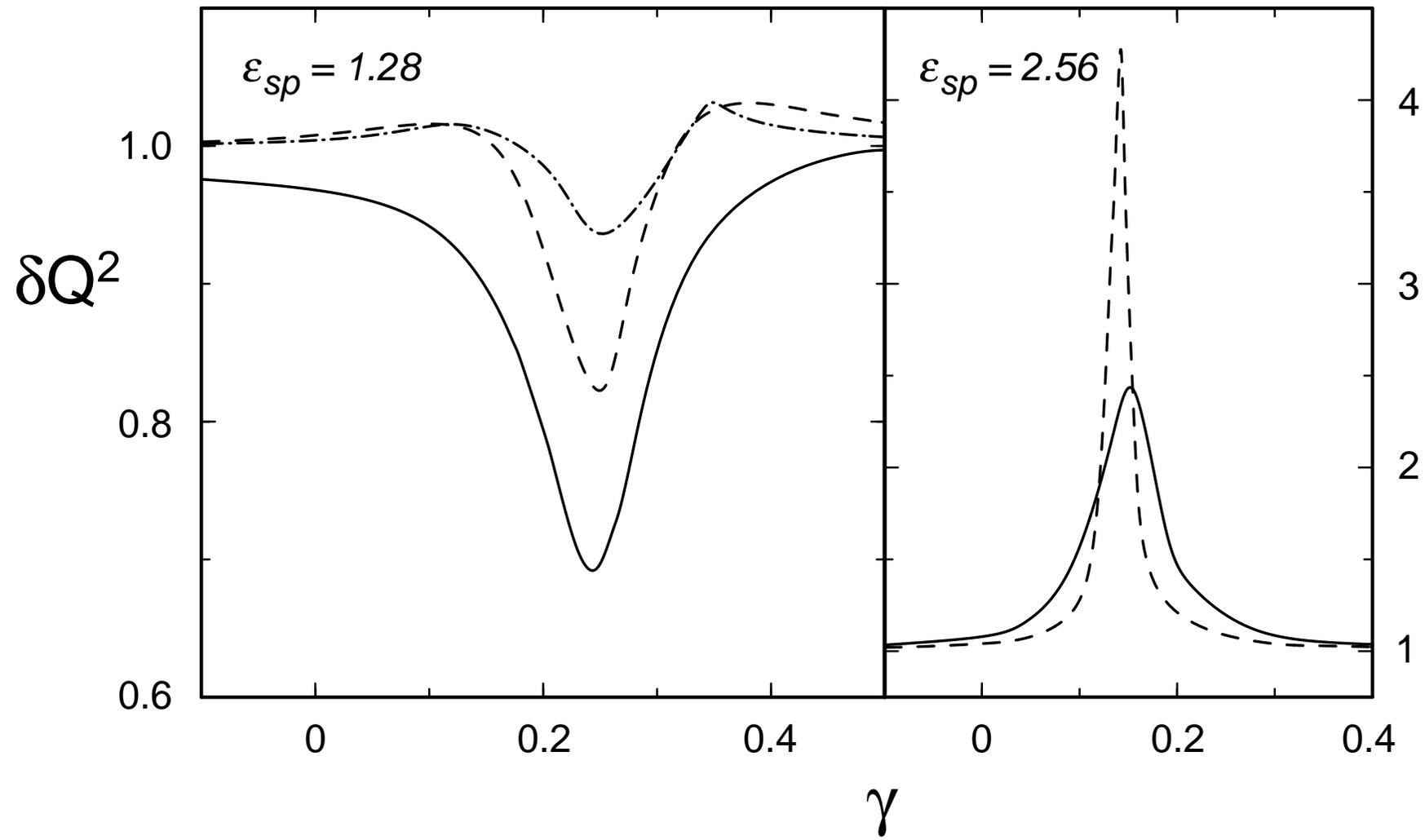
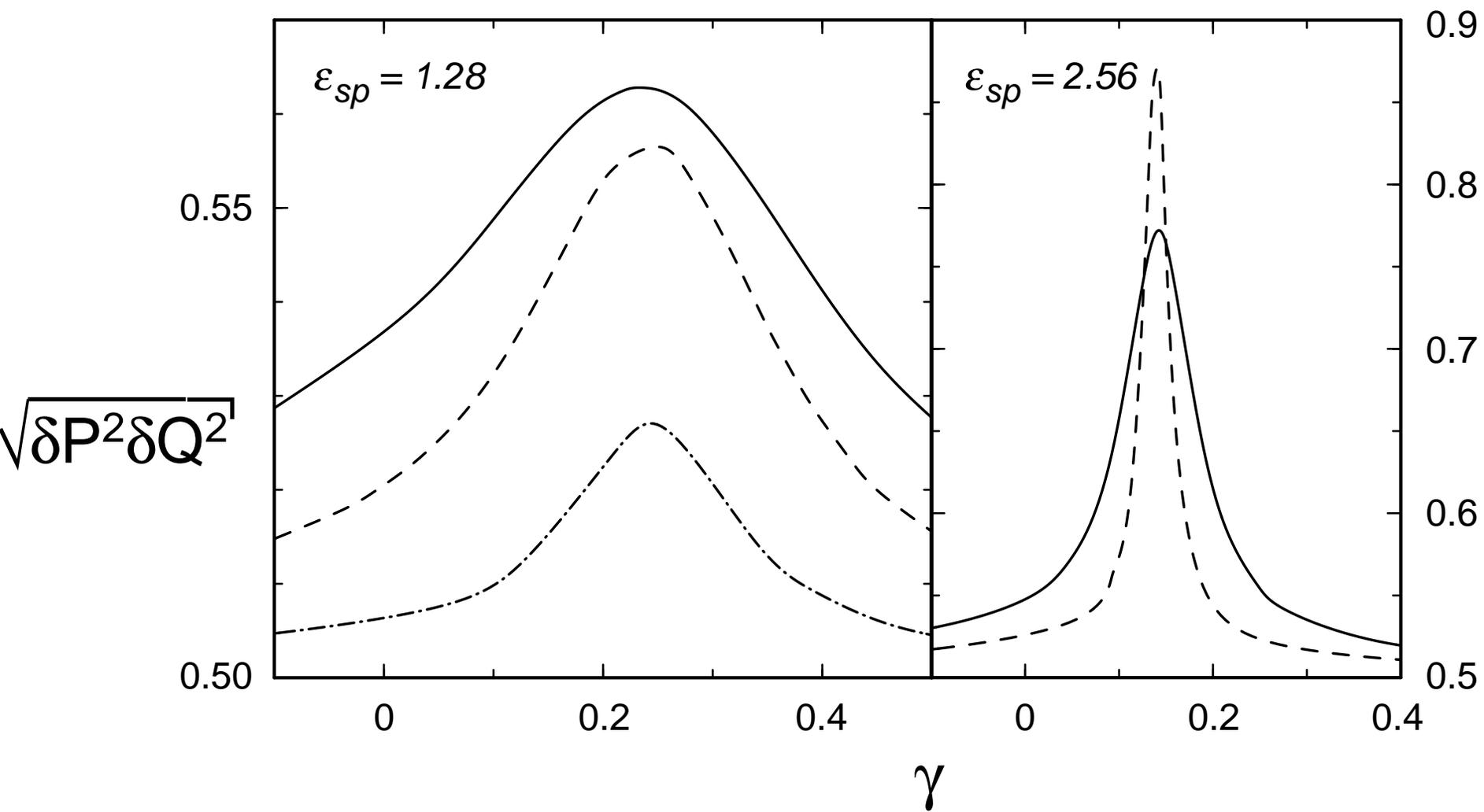
\end{document}